\documentclass{svjour2}                    
\smartqed  

\usepackage[latin1]{inputenc}

\usepackage{amsfonts}
\usepackage{amsmath,amssymb}
\usepackage{graphicx}

\usepackage{array}

\newlength{\defbaselineskip} 
\newcommand{\setlinespacing}[1]%
           {\setlength{\baselineskip}{#1 \defbaselineskip}} 
\newcommand{\doublespacing}{\setlength{\baselineskip}
                           {2.0 \defbaselineskip}}

\usepackage{lineno}

\fontsize{12}{15}

\newlength{\longtitre} 
\newcommand{\au}{\bar{u}}
\newcommand{\av}{\bar{v}}
\newcommand{\ax}{\bar{x}}
\newcommand{\ay}{\bar{y}}
\newcommand{\ap}{\bar{p}}

\newcommand{\partialx}[1]{\partial_{\ax} #1}

\newcommand{\partialy}[1]{\partial_{\ay} #1}

\newcommand{\secondy}[1]{\partial_{\ay^2}^2 #1}

\newcommand{\vecteur}[1]{\mathsf{#1}}

\newcommand{\tenseur}[1]{\boldsymbol{#1}}

\newcommand{\identity}{\tenseur{I}}

\newcommand{\smalldeftensor}{\tenseur{\varepsilon}}

\newcommand{\stresstensor}{\tenseur{\sigma}}

\newcommand{\young}{E}

\newcommand{\poisson}{\nu}

\journalname{
Medical \& Biological Engineering \& Computing
}

\begin{document}

\title{
Modelling the human pharyngeal airway:
}

\subtitle{
validation of numerical simulations using in-vitro experiments
}


\author{
Franz Chouly 
\and Annemie Van Hirtum 
\and Pierre-Yves Lagrée 
\and Xavier Pelorson 
\and Yohan Payan 
}


\institute{
Franz Chouly \at
INRIA, REO team, Rocquencourt - BP 105, 78153 Le Chesnay Cedex, France.\\
Tel.: +33 (0)1 3963 5916\\
Fax.: +33 (0)1 3963 5882\\
\email{franz.chouly@inria.fr}\\\\
Yohan Payan \at
Laboratoire TIMC, UMR CNRS 5525, Université Joseph Fourier,\\
38706 La Tronche, France.\\
\email{yohan.payan@imag.fr}\\\\
Annemie Van Hirtum \& Xavier Pelorson \at
Département Parole et Cognition, GIPSA-lab,
INPG / UMR CNRS 5216,\\
46 Av. Felix Viallet, 
38031 Grenoble, France.\\
\email{annemie.vanhirtum@gipsa-lab.inpg.fr / xavier.pelorson@gipsa-lab.inpg.fr}\\\\
Pierre-Yves Lagrée \at
Institut Jean le Rond d'Alembert,
UMR CNRS 7190,\\
4, Place Jussieu, 
75252 Paris Cedex 05, France.\\
\email{pierre-yves.lagree@upmc.fr}
}

\date{Received: 09/05/08 / Accepted: 01/10/08\\
{
}
}

\maketitle


\begin{abstract}
In the presented study, a numerical model which predicts the 
flow-induced collapse within the
pharyngeal airway is validated using in-vitro measurements.
Theoretical simplifications were considered to limit the computation time.
Systematic comparisons between simulations and measurements were performed on an 
in-vitro replica, which reflects asymmetries of the geometry and of 
the tissue properties at the base of the tongue and in pathological conditions
(strong initial obstruction).
First, partial obstruction is observed and predicted.
Moreover, the prediction accuracy 
of the numerical model
is of 4.2 \% concerning
the deformation (mean quadratic error on the constriction area).
It shows the ability of the assumptions and method
to predict accurately and quickly a fluid-structure interaction.

\end{abstract}

\section{Introduction}

Since the 90's, biomechanical modelling of the human upper airway has
received a growing interest since it allows a better understanding of its
physiology and pathophysiology, as well as an increased quality of the treatments
of its specific pathologies.
Among these, Obstructive Sleep Apnea Syndrome (OSAS) has been
object of particular attention, as it became a major
health care topic, affecting a growing part of the population, especifically in Europe
and in the United States \cite{malhotra02a,young93a}.
It is characterized by the occurence of an abnormal rate
of apneas and hypopneas during sleep \cite{malhotra02a}. 
During an episode of obstructive apnea (respectively hypopnea), the soft tissue in the pharynx 
completely (respectively partially) collapses in response to inspiratory airflow. It induces a
temporary cessation (respectively limitation) of the respiration \cite{ayappa03a}. 
The main health effects are excessive daytime sleepiness and an increased
risk of cardiovascular diseases \cite{malhotra02a}.
In parallel to the great amount of medical/biomedical engineering research 
that has been carried out to
understand this highly complex phenomenon 
(see e.g. 
\cite{yamashiro07a,abeyratne07a} for recent works
or \cite{malhotra02a,ayappa03a,fogel04a,fairbanks03a,ferber07a} 
for recent overviews), 
theoretical biomechanical models have
been proposed such as 
simplified mathematical models of the interaction between the airflow and
the soft tissue
\cite{gavriely93a,fodil97a,aittokallio01a},
numerical models of the respiratory fluid flow
\cite{shome98
,liu07a}
and numerical models of the fluid-structure interaction
\cite{payan01a,malhotra02b,xu05a}.
They allow indeed to understand the relationship between the 
fluid-structure interaction in the pharynx
and the abnormal flow patterns observed in the
medical or physiological experiments. They might allow furthermore to
predict these flow patterns using measurements of the biomechanical 
properties of the upper airway (geometry, rheology). This makes them of
interest to improve the quality of the treatments such as surgical
procedures \cite{crampette92c,guilleminault89a}
or mandibular advancement splints \cite{hui00a}. 
\\

In a previous paper \cite{choulyXXa}, a numerical model which aims
at predicting the fluid-structure interaction within the pharyngeal airway
has been described. It is based on simplified assumptions so that
the computational cost of simulations be compatible with
clinical applications. 
An original in-vitro tongue replica has also been described. It
takes into account the very specific properties of the upper airway.
In particular, the asymmetry of the pharyngeal duct at this level has
been reproduced in a simplified manner.
A few comparisons between predictions of the numerical model
and measurements of the flow-induced collapse within the in-vitro replica
provided encouraging results (see \cite{choulyXXa} for details).
Nevertheless, this was not sufficent to be really
conclusive about the interest and drawbacks of the numerical model.
Therefore, the aim of this study was 
extensive validation of this numerical model through
in-vitro experiments, and to our
knowledge, it has not been done so far in this context (for
instance in \cite{xu05a}, the validation is carried out with
rigid walls).
The main novelty in comparison to the study presented in 
\cite{choulyXXa} is that 
systematic comparisons have been carried out, for a wide
range of parameter values. 
Moreover, pressure sensors have been integrated
into the experimental setup. As the pressure at the level of the
constriction plays a critical role in the flow-induced collapse
(see \cite{vanhirtum04a}), it was jugded interesting to assess
its prediction in conditions of fluid-structure interaction. 
Besides, a digital camera quantifies the two-dimensional
deformation of the simplified airway at the main site of obstruction.
This gives more detailed and pertinent information than
the laser beam which was of use in \cite{choulyXXa}.\\

\section{Material and methods}
\label{sect:theory}

\subsection{In-vitro setup}
\label{sub:invitrosetup}

The
aim of the experimental setup described here is to reproduce a fluid-structure interaction
in conditions 
of strong obstruction 
and quasi-steady motion.
It obviously simplifies the complex in-vivo reality,
but allows
to obtain reliable dynamical measurements of the pressure and of the deformation
in controlled conditions, 
which is required for validation.\\

\begin{figure}[t]
\center

\vspace{2cm}

\begin{minipage}[t]{8.1cm}
\includegraphics[angle=0,width = 8.0cm]
                {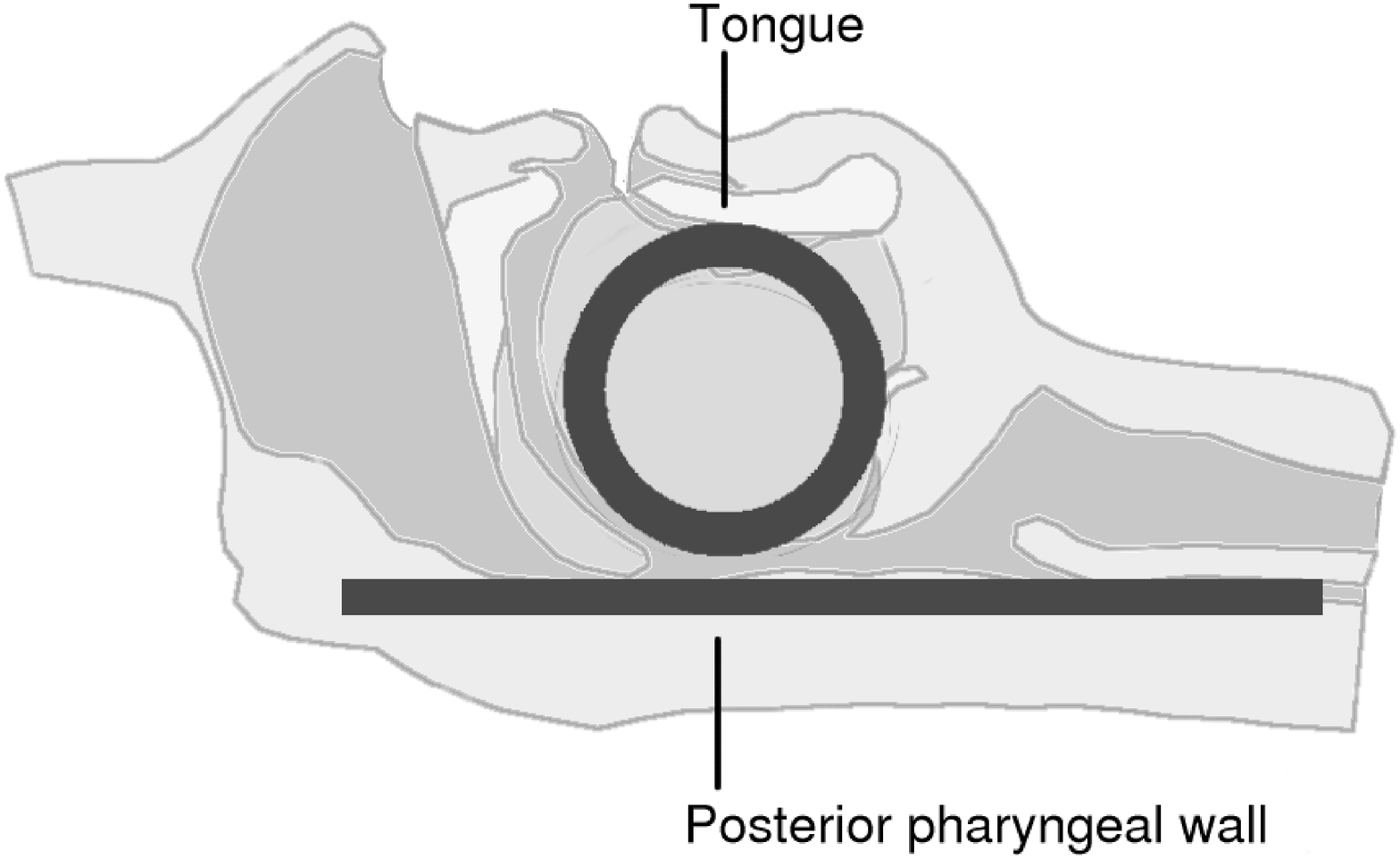}
(a)
\end{minipage}
\begin{minipage}[t]{8.1cm}
\includegraphics[angle=0,width = 8.0cm]
                {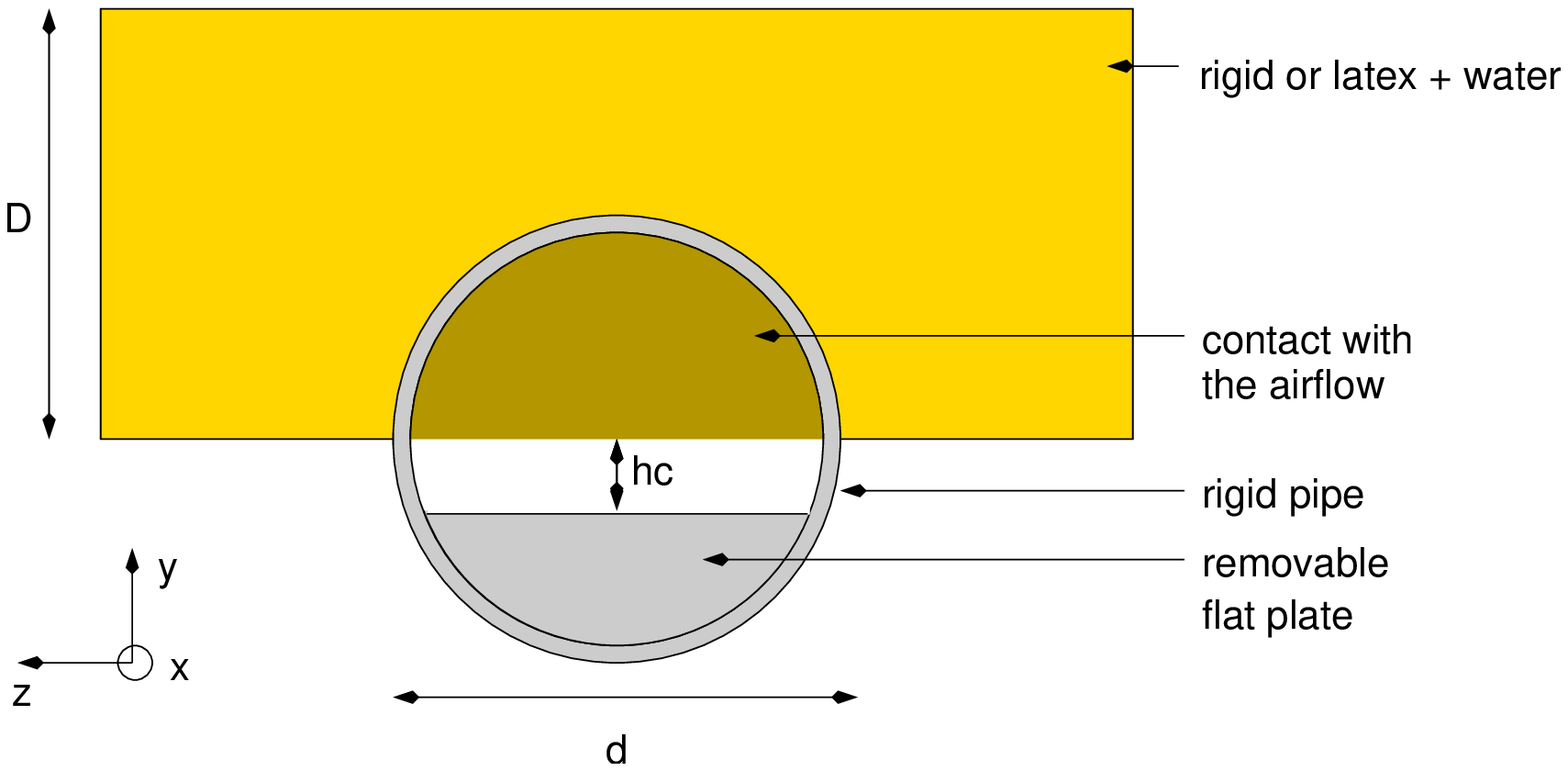}
(b) 
\vspace{2mm}
\end{minipage}
\begin{minipage}[t]{15.2cm}
\includegraphics[angle=0,width = 15.1cm]
{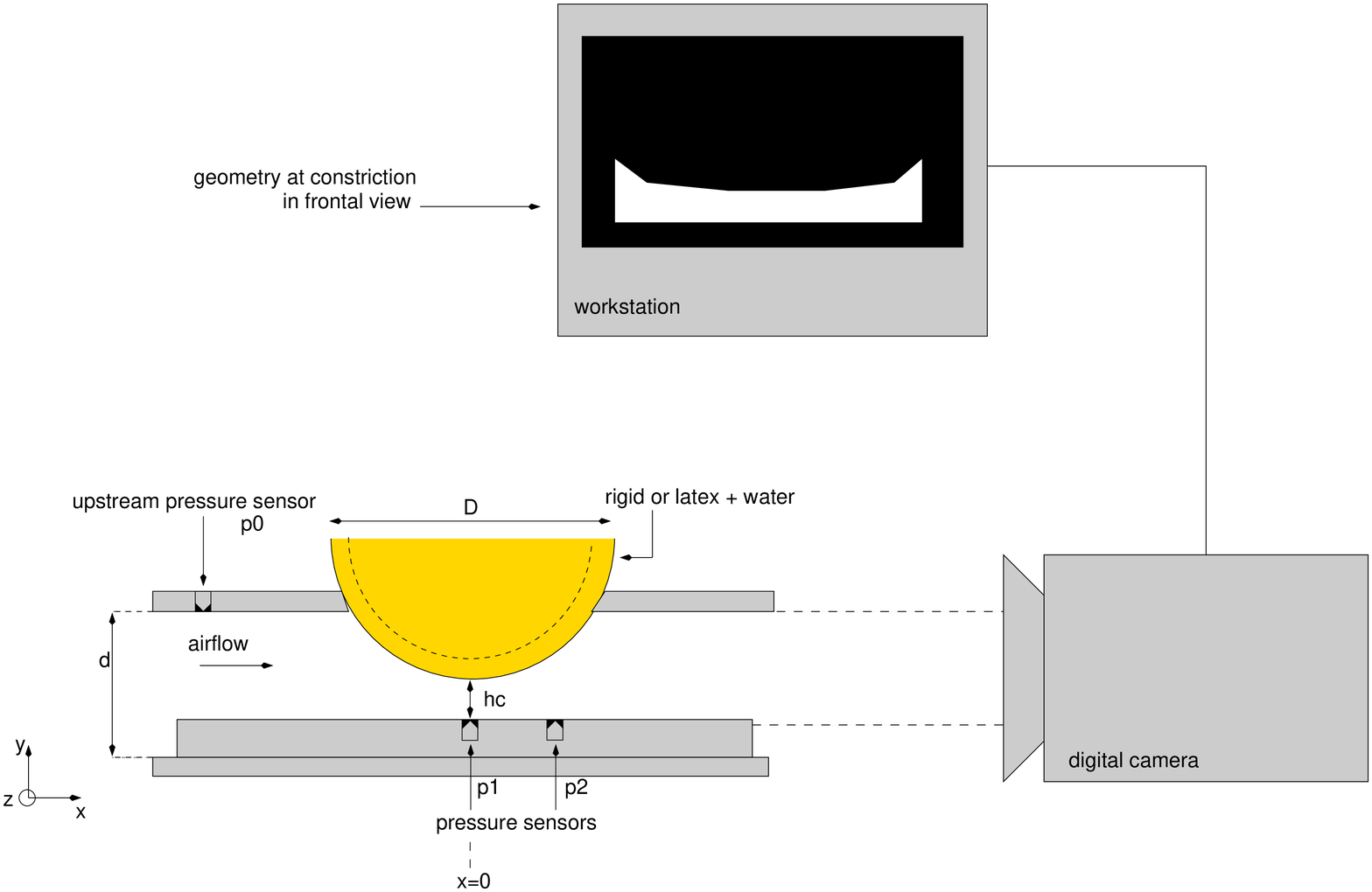}
(c) 
\end{minipage}

\caption{
Sagittal view of the upper airway overlaid with
an outline of the in-vitro setup (a).
Outline 
of the in-vitro setup, from
frontal view (b).
Experimental measurement of the wall deformation with a digital
camera (c). 
The acquisitions of the camera are treated automatically
with a software. Pressure sensors 
allow to obtain at each time step information about the airflow, which complements
the information that comes from the camera.
} 

\label{img:invitrosetup}
\end{figure}

\begin{figure}[t]
\center

\vspace{5cm}

\begin{minipage}[t]{7.5cm}
\includegraphics[angle=0,width = 7.4cm]
                {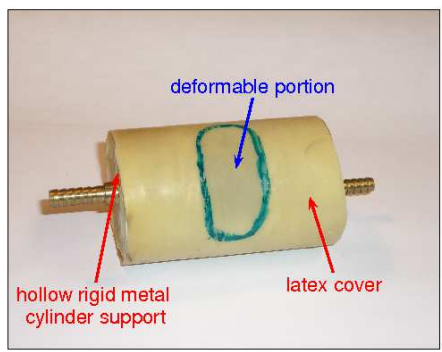}
(a) 
\end{minipage}
\begin{minipage}[t]{7.5cm}
\includegraphics[angle=0,width = 7.4cm]
                {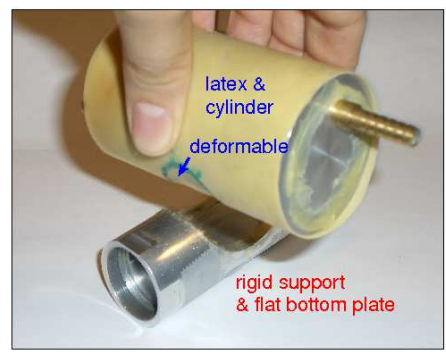}
(b) 
\end{minipage}
\begin{minipage}[t]{7.5cm}
\includegraphics[angle=0,width = 7.4cm]
                {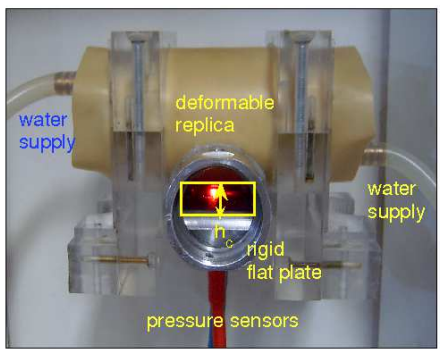}
(c) 
\end{minipage}

\caption{
Photographs of the in-vitro setup.
(a) Deformable upper cylinder: view of the deformable portion.
(b) The mounting of the 'in-vitro' tongue replica.
(c) Frontal view of the replica at downstream end.
} 

\label{img:invitrosetup-photos}
\end{figure}

The in-vitro setup is depicted Figures \ref{img:invitrosetup} \&
\ref{img:invitrosetup-photos}.
The real morphology of the upper airway
has been simplified as suggested Figure \ref{img:invitrosetup} (a).
However, its overall configuration and the main dimensions have been conserved. 
In particular, a frequent site
of 
collapse is the base of the tongue \cite{rama02a}.
This anatomical entity is represented in the setup by a short latex tube filled with
water (deformable tongue replica). This deformable structure
intersects orthogonally a rigid metallic pipe which stands for the pharynx.
The geometry of the duct at the level of this junction is those of
a constriction, in 
accordance with observations at the tongue base.\\

The diameter $D$ of the deformable tongue replica is 49 mm.
The latex wall (Piercan Ltd.) has a thickness $e$ of 0.3 mm.
Its Poisson's ratio $\poisson$ is of 0.5 (incompressible material). 
The value of its Young modulus $E$ has been determined to 1.68 MPa 
(see \cite{chouly05a} for details).
The latex sheet is mounted onto a rigid metallic support (Figure 
\ref{img:invitrosetup-photos} (a)). 
A hole in the support enables the deformation of the latex in response to the fluid flow:
the deformable part of the
cylinder is  depicted Figures \ref{img:invitrosetup-photos} (a) 
\&
\ref{img:invitrosetup} (b) 
in dark shade.
An external water supply (Figure \ref{img:invitrosetup-photos} (c)) 
through
a duct connected to a water column allows
to control and to measure the water pressure $P_{ext}$ inside the latex 
interface (external flow). Indeed, the height of the water column is controllable
\cite{vanhirtum05a}.
Manipulation of this pressure influences simultaneously two parameters:
the geometry of the duct and the overall stiffness of the deformable part.\\

The rigid metallic pipe in which the airflow (internal flow) circulates
is depicted Figure \ref{img:invitrosetup-photos} (b). 
Its diameter $d$ is
of 25 mm. A removable flat plate allows to change the geometrical characteristics
of the duct 
(Figure \ref{img:invitrosetup} (b)).
Different heights\footnote{dimension in $y$ direction} of the duct,
corresponding to different pharyngeal calibers, can be chosen using different
plates. 
Air supply comes from a pressure reservoir, which can be considered
as an artificial lung. This is a rectangular box of 
approximatively $0.75$ $\mathrm{m^{3}}$, fed by a compressor.
The control of the pressure in the reservoir is ensured by a pressure
regulator (Norgren(TM) type 11-818-987). 
A constant selected pressure, within the range 0-3000 Pa, is thus
obtained at the inlet of the duct
\cite{ruty07a}.
The pressure within the airflow is measured with sensors located at different
positions 
(Figure \ref{img:invitrosetup} (c))
One sensor ($p_{0}$) allows
the measurement of the upstream pressure. Two others ($p_{1}, p_{2}$)
are located at the bottom of the rigid plate, and allow measurements of the 
pressure at the level of the constriction and downstream. 
The pressure sensors are within holes
of diameter 0.4 mm. They are piezo-electric\footnote{{\it Endevco 8507C,
Kulite XCS-093}} and deliver a tension proportional to the pressure.
Preliminary calibration, with a liquid manometer (Kimo(TM)), allows to determine the
constant of proportionality of the relationship. Then, the pressure is measured
with a precision of $\pm$ 1 mm$\mathrm{H_{2}O}$.\\

An important geometrical parameter is the minimal height of the duct, 
at the bottom of the upper cylinder, called the constriction height $h_{c}$
(Figures \ref{img:invitrosetup} (b) (c) \& \ref{img:invitrosetup-photos} (c)). 
As explained before, this height could be changed using different metallic plates, while
$P_{ext}$ is maintained.
In physiological terms, it corresponds to the minimal pharyngeal caliber.
Another important parameter is the area at the level of the constriction $A_{c}$. 
It corresponds to the minimal pharyngeal area.
We chose this
parameter for the measurement of the flow-induced deformation
since it captures its global effect on the obstruction of the duct
($h_{c}$ is only a punctual measurement). 
To this purpose,
a digital camera has
been used (Figure \ref{img:invitrosetup} (c)), 
which measures the bidimensional geometry at the level of the constriction,
in frontal view and at the downstream end (Figures \ref{img:invitrosetup} (b) \&
\ref{img:invitrosetup-photos} (c)). The resulting information is the constriction
height $h_{c}(z)$ for all $z$, from which we
obtain the constriction area $A_{c}$
and 
follow its variation.\\
 
The digital camera is an industrial model\footnote{{\it Inca 311, Philips}},
with a resolution of $1280 \times 1024$ pixels. 
Its focal has been adjusted so as to visualize with the best
accuracy the constriction site (focal plane located at the site of the constriction,
and as wide as the rigid pipe). The zoom objective is such that the
number of pixels corresponding to the constriction area is maximized.
The contrast is enhanced with an ordinary light source upstream the pipe. 
A software\footnote{{\it Clicks(TM), version 1.4.0, 
Philips, Industrial Vision}} allows to
control the parameters of the camera, to visualize and to store the resulting acquisitions.
Those can be obtained automatically, at periodic time intervals, in synchronization 
with the pressure measurements, thanks to LabView\footnote{LabView 7, National Instruments}. 
A preliminary calibration is necessary, to determine the relationship between
distances measured on the picture and real distances. The method consists in localizing
three points on the picture. The distance between these points is known. 
After the calibration, the horizontal
and vertical resolutions are determined: $r_{h} = r_{v} \simeq 6.3.10^{-3}$ cm / pixel.\\

\subsection{Mechanical theory for fluid-structure interaction}

Numerical computation of a complete fluid-structure interaction problem, i.e.
solving the incompressible Navier-Stokes equations dynamically coupled with 
a deformable structure in large deformations, is still challenging nowadays,
both in terms of stability and convergence towards the solution and 
in terms of computation
cost \cite{shome98,li05a,tada05a,wolters05a}. 
Therefore,  
the complexity of the
description of the ongoing phenomena is reduced, and simplificatory
assumptions are stated.
The further outlined assumptions are made in agreement with the anatomy and the physiology
of the upper airway, and they allow 
to reduce the computation time to a great extent.
A detailled description of the biomechanical model 
and of the numerical method can be found in \cite{choulyXXa} and 
\cite{chouly05a}. In this section, only the main characteristics are given.\\

Concerning the structure, 
the assumptions of
a quasi-steady motion, with small deformations and displacements
have been chosen. The constitutive behavior is given by the Hooke
law (linear elasticity).
As a result, the following equations govern the structure deformation:

\begin{equation}
\label{eqn:equilibrestatique}
\left \{
\begin{array}{l}
\nabla \cdot \stresstensor + \vecteur{f} = \vecteur{0},\\\\

\stresstensor = 
\frac{\young \poisson}{(1 - 2 \poisson) (1 + \poisson)} \: 
\smalldeftensor_{I} \identity
+ \frac{\young}{1 + \poisson} \:
\smalldeftensor,\\\\

\smalldeftensor
=
\frac{1}{2} ( \nabla \vecteur{u} + (\nabla \vecteur{u})^t ),
\end{array}
\right.
\end{equation}

where $\stresstensor$ is the Cauchy stress tensor, and $\vecteur{f}$
is the vector of external forces. 
$\young$ is the Young's modulus, 
and $\nu$ is the Poisson's ratio. 
$\smalldeftensor_{I}$ is the trace (first invariant) of the small deformation tensor
$\smalldeftensor$.
$\vecteur{u}$ is the vector of displacement. 
The boundary conditions consist in
immobility constraints (displacement $\vecteur{u} = 0$) at sites of
attachement to rigid parts 
and imposed external forces in the surface 
of contact with the airflow.\\

The fluid flow is considered as incompressible (Mach number of $O(10^{-2})$),
laminar (Reynolds number of $O(10^{3})$), stationnary (Strouhal number of
$O(10^{-3})$) and bidimensional \cite{vanhirtum04a}. 
Since the Reynolds number is of the order of $10^3$, we use an asymptotic
simplification of the incompressible Navier-Stokes equations:

\begin{equation}
\label{rnsp_equations}
\left \{
\begin{array}{rcl}
        \au \partialx{\au} + \av \partialy{\au} & = & - \partialx{\ap}
                                                      + \secondy{\au},\\\\
        -\partialy{\ap} & = & 0,\\\\
        \partialx{\au} + \partialy{\av} & = & 0,\\
\end{array}
\right.\\\\
\end{equation}

$(\au,\av)$ are the longitudinal ($\ax$) and transverse ($\ay$) components of the
fluid velocity, and $\ap$ is the pressure \cite{lagree05b}.
All the variables are nondimensional:
$\ax = x (h_0 Re)^{-1}$,
$\ay = y h_0^{-1}$,
$\ap = P (\rho U_0^2)^{-1}$,
$\au = u U_0^{-1}$,
$\av = v Re U_0^{-1}$;
where $h_0$ is the transversal dimension of the pharyngeal duct, $U_0$ is the mean longitudinal
speed, and $Re$ is the Reynolds number ($Re = U_0 h_0 / \nu $, with $\nu$ the
kinematic viscosity of the air).
These equations, called Reduced Navier-Stokes / Prandtl (RNSP), 
allow to take into account the boundary layer formation
as well as the separation of the fluid after the narrowing of the pharyngeal duct, at the
base of the tongue \cite{lagree05b}. The boundary conditions are the following:
no slip on the upper and lower walls and a pressure difference $\Delta P$ imposed
between the inlet and the outlet.\\

\subsection{Numerical solving of the equations}

As the problem is considered as quasi-steady, the fluid-structure interaction 
is solved using a segregative method: the equations that govern the fluid and 
the solid are solved alternatively, inside a global loop. The pressure gradient in the
fluid is imposed gradually, in $st$ steps : $\Delta P_{1} = 0, \ldots , 
\Delta P_{st} = \Delta P_{max}$. For each pressure gradient $\Delta P_{i}$
at step $i$, the fluid forces on the wall are first computed and imposed. Then,
the wall is deformed, which changes the fluid flow domain. The fluid forces
need then to be computed again. This is the reason why
a finite number of iterations $it$ should be imposed at each step,
until equilibrium of the wall is reached. A convergence criterion $cv$,
which corresponds to the maximal
displacement between two iterations, allows
to ensure that at the end of the $it$ iteration, the displacement of 
the structure 
is no more significant. Typically, a choice of $st = 8$ and $it = 6$ has been
found sufficient to ensure the convergence of the algorithm ($cv \simeq
\mathrm{10^{-3}}$ mm).\\

\begin{figure}[t]
\center

\vspace{8cm}

\begin{minipage}[t]{7.5cm}
\includegraphics[angle=0,width = 7.4cm]
                {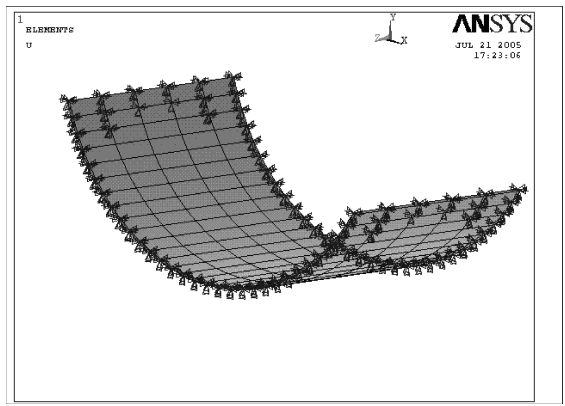}
\end{minipage}

\caption{
Finite element model of the latex wall of the in-vitro setup,
with boundary conditions. The nodes indicated by black arrows 
are immobilized, as they are supposed to be fixed to the rigid pipe.
}

\label{img:fem-bc}
\end{figure}

The continuum equations of the wall are solved using the finite element 
method \cite{zienkiewicz89a}. In the context of small deformations and
of linear elasticity, the relationship between the nodal displacements and
the nodal forces is linear. As a result, the precomputation of the inverse of the stiffness
matrix $\mathbf{[K]}$ is done before
the fluid-structure interaction
loop \cite{cotin99a}. 
It saves computation time as nodal displacements are obtained through
simple matrix multiplication at each iteration. 
This preliminary step is achieved using a commercial finite element solver
(Ansys(TM) Software).
The finite element model of the tongue replica, in other terms of the
latex tube filled with water, is depicted
Figure 
\ref{img:fem-bc}. 
The mesh is constituted of 150 linear elements of
8 nodes, regularly dispatched. The nodes in contact with the hollow
metal support (Figure \ref{img:invitrosetup-photos} (a)) are
immobilized (Figure \ref{img:fem-bc}). 
The boundary conditions are such
that a bidimensional model would have not been satisfying. This is
the reason for the choice of a tridimensional model. The incompressibility
of the latex is approximated through the choice of a Poisson's ratio $\poisson$
of 0.499. 
The Young modulus has been fixed to $E=1.68$ MPa.
The water inside the latex tube has been taken into account in the model
through the application of constant pressure forces on the surfaces supposed
to be in contact with the water.\\
 
The RNSP equations that govern the fluid flow are solved using a finite
difference method \cite{lagree05a}, which is easy to implement and
allows fast numerical solving. 
The grid that is used is of size 
$2000 \times 1000$.
This is sufficently high to capture boundary layer formation and separation,
as well as jet formation, and increasing the grid resolution above this level
does not significantly change the values of the pressure distribution
(see \cite{chouly05a} and \cite{choulyXXa} for details).
As the airflow is modelled using bidimensional equations, the finite element model is
divided into 5 thin slices on which are carried out computations of the fluid forces.
A finite element model of 10 slices has also been built so as to check the impact
of this numerical parameter and no significant difference between the two
models has been found.
The resulting pressure distribution is imposed at the surface of the structure
in agreement with the principle of virtual works, which provides an accurate
approximation of the pressure load (see \cite{choulyXXa} for detailed formulas). 
The codes for computation of the fluid flow and fluid-structure interaction have 
been written by the authors, using Matlab(TM) interfaced with C language.
As a result, the duration of the computations for the overall algorithm
is typically of the order of 20 minutes\footnote{Dell Precision 330 (TM) 
workstation, with Pentium(TM) 4, 2 GHz, 1 GO RAM.
}. 
This can be considered as
reasonable for clinical applications.

\subsection{Methodology for comparisons}
\label{sub:methodology}

A batch of measurements has been carried out, for which the
latex thickness ($e = 0.3$ mm) and the initial constriction height
($h_{c}^{0} = 1.84$ mm)\footnote{at $P_{ext} = 0$ Pa}
remained fix. 
The following parameters were
systematically varied:

\begin{itemize}

\item  the external pressure $P_{ext}$,

\item the maximal inlet pressure $P_{e}^{max}$.

\end{itemize}

The external pressure ranges from 100 to 700 Pa. For each value
of $P_{ext}$, four measurements have been carried out, corresponding to
different values of $P_{e}^{max}$ : 200, 400, 600 and 800 Pa. 
Measurements are performed according to the following procedure: 
the inlet pressure $P_{e}$ is gradually increased from 0 Pa 
up to the required value $P_{e}^{max}$. At fixed time
steps (0.25 s), a synchronized sampling of the following outputs is done automatically:

\begin{itemize}

\item the upstream pressure $P_{e}$, the pressure $P_{c}^{0}$ at $x=0$ mm, 
and the pressure $P_{c}^{16}$ at $x=16$ mm, as measured respectively by
the sensors $p_{0}, p_{1}, p_{2}$ described in Section \ref{sub:invitrosetup}. 
Note that when $P_{e} = 0$ Pa, $P_{c}^{0}$ corresponds to the pressure at
the level of the constriction.
Yet, during the deformation, the new position of the wall
can be such that the pressure $P_{c}^{0}$ does no longer correspond to
the pressure at the constriction.

\item the height of the duct at the level of the constriction : $h_{c}(z)$, using
the digital camera described in Section \ref{sub:invitrosetup}.
The constriction area $A_{c}$ is determined automatically. 

\end{itemize}

The experiment is stopped when the value observed for the output does not
change anymore. The stationnary state is then reached, after approximatively
10 seconds.\\

For each value of $P_{ext}$, the mean experimental curves 
$P_{c}^{0,exp} (P_{e})$ 
and $A_{c}^{exp} (P_{e})$
that result from 
the four experiments at different values of $P_{e}^{max}$ are computed, with the
associated standard deviation. A numerical simulation is then carried out.
The inlet pressure in the simulation is equal to 
$P_{e}$ and the
outlet pressure is equal to $P_{c}^{16}$. 
The theoretical curves 
$P_{c}^{0,sim} (P_{e})$ 
and $A_{c}^{sim} (P_{e})$ are compared to the experimental ones: 
$P_{c}^{0,exp} (P_{e})$ 
and $A_{c}^{exp} (P_{e})$. 
For quantitative comparison 
we compute first the mean quadratic error, in normalized form, which is,
for the constriction area:

\begin{equation}
\label{eqn:erreur2}
\tilde{\varepsilon}_{2} (A) = 
\frac{1}{A_{c}^{P^{ext}}}
\left [ 
\frac{1}{n}
\sum_{i=1}^{n} 
\left (
A_{c}^{exp} (P_{e}^{i}) - A_{c}^{sim} (P_{e}^{i})
\right )^{2}
\right ]^{\frac{1}{2}}.
\end{equation}

Here, $n$ is the number of points in the experimental sample,
$(P_{e}^{i}, A_{c}^{exp}(P_{e}^{i}))$ the coordinates of the $i^{th}$
experimental point, $(P_{e}^{i}, A_{c}^{sim}(P_{e}^{i}))$ the coordinates
of the corresponding point on the theoretical curve. 
The constant of normalization $A_{c}^{P^{ext}}$ is
the area at the beginning of the experiment, when no internal airflow
is circulating ($P_{e} = 0$ Pa) and after a water pressure
$P_{ext}$ has been imposed. For the pressure $P_{c}^{0}$, the quadratic
error $\tilde{\varepsilon} (P^{0})$ is computed identically, except
the constant of normalization which is the maximal
inlet pressure: $P^{max}_{e}$.
A maximal error has been
computed as well:

\begin{equation}
\label{eqn:erreurmax}
\tilde{\varepsilon}_{max} (A) = 
\frac{1}{A_{c}^{Pext}}
\max_{1 \leq i \leq n}
\left | 
A_{c}^{exp} (P_{e}^{i}) - A_{c}^{sim} (P_{e}^{i})
\right |,
\end{equation}

(and an identical definition for $\tilde{\varepsilon}_{max} (P^{0})$).
Finally, the overall prediction performance is evaluated using the coefficient of
determination $R^{2}$ ($0 \leq R^{2} \leq 1$):

\begin{equation}
R^{2} = 1 - \frac{\hat{\sigma}^{2}}{\sigma_{y}^{2}},
\end{equation}

$\hat{\sigma}^{2}$ being the variance of the prediction residuals,
and $\sigma_{y}^{2}$ is the variance of the experimental measurements.
The closer is $R^{2}$
to 1, the best is the prediction of the numerical simulation. 
As a result, two
coefficients of determination are computed: 
$R_{P}^{2}$ for the pressure at the level of the (initial) constriction $x=0$
and $R_{A}^{2}$ for the constriction
area.\\

\section{Results}
\label{sect:results}

\begin{figure}[t]
\center

\vspace{3cm}

\begin{minipage}[t]{10.2cm}
\includegraphics[angle=0,width = 10.1cm]
{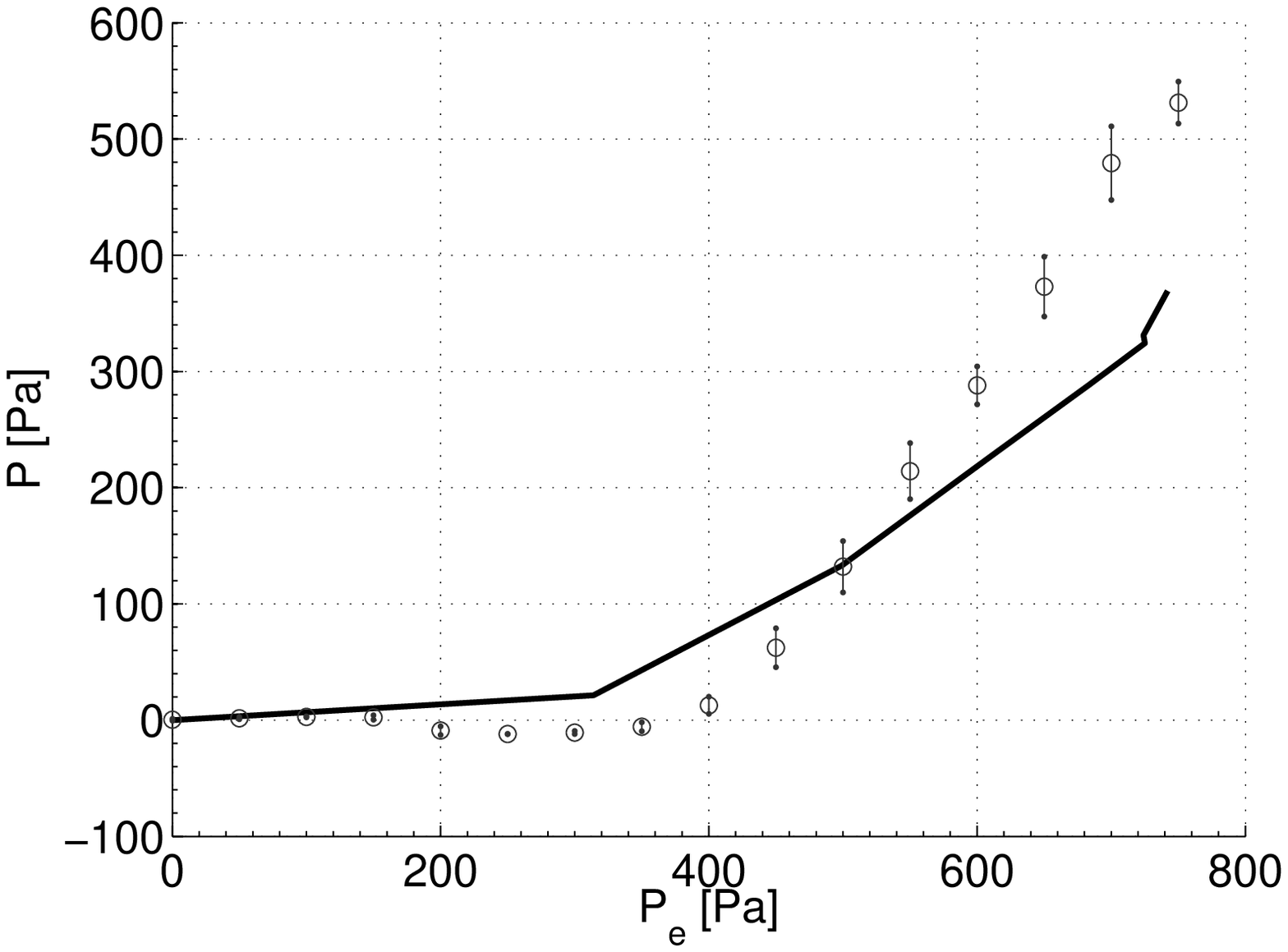}
(a) 
\end{minipage}

\begin{minipage}[t]{10.2cm}
\includegraphics[angle=0,width = 10.1cm]
{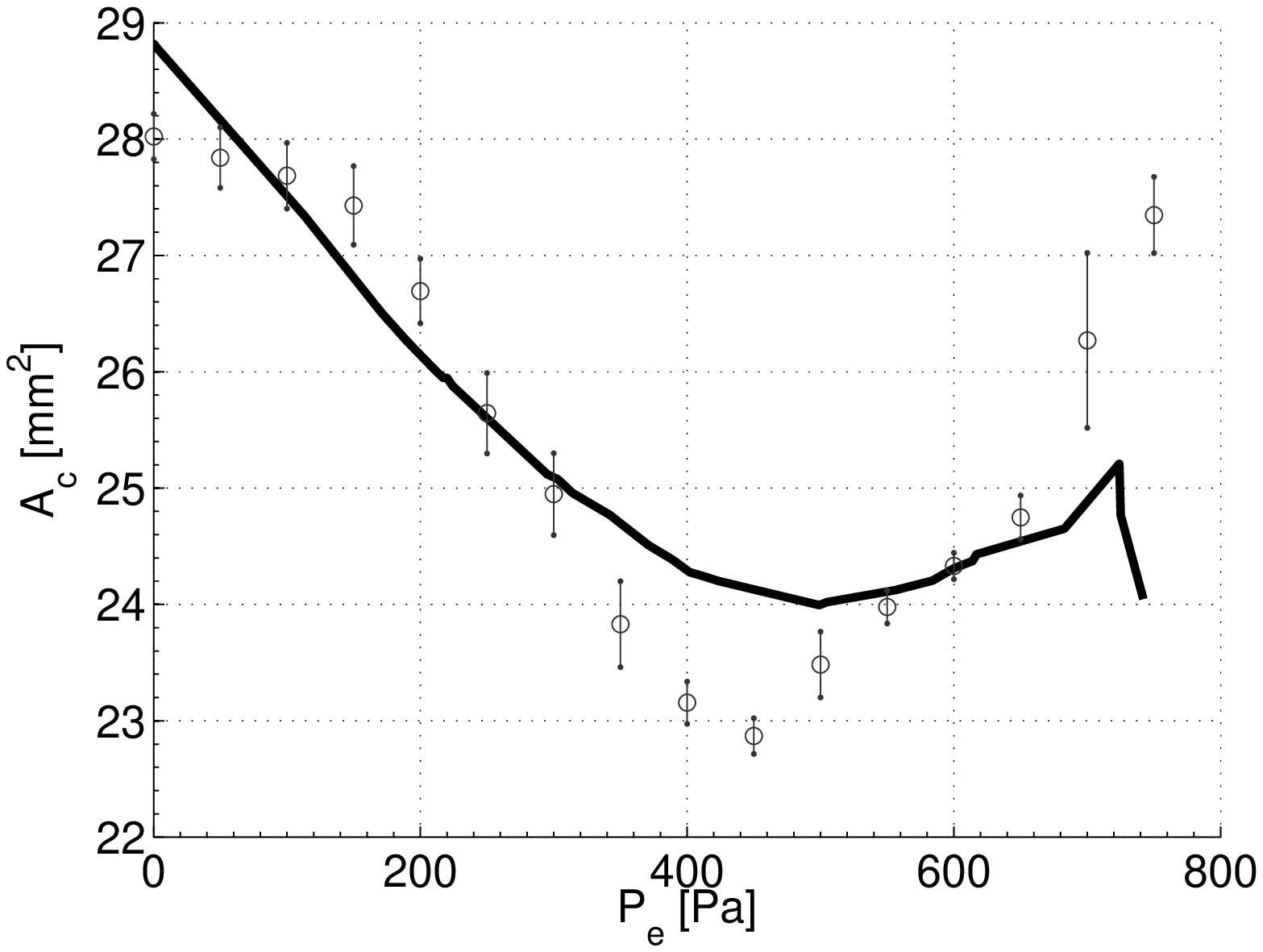}
(b) 
\end{minipage}

\caption{
Example of a comparison between the measurements (mean value
with standard deviation, in grey) and the simulations (black curve)
for 
$P_{ext} = 300$ Pa.
The pressure $P_{c}^{0}$ (a) and the 
constriction area $A_{c}$ (b) are compared.
}

\label{img:cmp100-200}
\end{figure}

The results of the comparisons between predictions and measurements
are summarized Table \ref{tbl:cmpError}, for all the investigated
values of $P_{ext}$, and a typical example of
the resulting curves $P_{c}^{0} (P_{e})$ 
and $A_{c} (P_{e})$ 
is given Figure \ref{img:cmp100-200}, for $P_{ext} = 300$ Pa.
First, it should be noticed 
that the measurements are strongly repeatable. Table \ref{tbl:cmpError}
indicates a mean value of the standard deviation $\sigma_{exp}$,
computed from the four repeated measurements, 
of 11.3 Pa for the pressure $P_{c}^{0}$ and of 0.29 $\textrm{mm}^{2}$
for the constriction area $A_{c}$. Variations of this standard deviation
with $P_{e}$ can be observed Figure \ref{img:cmp100-200}.\\

First, concerning the pressure $P_{c}^{0}$ at $x = 0$ mm, the theoretical and experimental
curves are in agreement with a mean error of less than 8 \%, until $P_{e}$ exceeds 500 Pa.
After this value, they split, 
and the value of
the computed pressure is always lower than the experimental value (of $\simeq$ 30
\%, see Figure \ref{img:cmp100-200} (a)). 
Moreover, this pressure is both negative and decreasing at the origin, but quickly
changes to become positive and increasing. This could be explained by the translation of the
constriction towards downstream as $P_{e}$ is increased and the wall is deformed. 
As a result, the pressure $P_{c}^{0}$ moves away from the value
of the pressure at the constriction, that remains still negative. The most valuable
explaination for the difference observed between the theoretical and the experimental
values of $P_{c}^{0}$ is that the simulation underestimates the displacement of the
wall in the $x$ direction. Considering the shape of the curve $P(x)$ in the area of the
constriction, an error of the order of 1 mm in the prediction of the displacement
should be followed by an error of the order of 100 Pa in the prediction of the 
pressure.
These phenomena lead to a mean quadratic error of 9.8 \% and
a mean value for $R^{2}_{P}$ of 0.64 (see Table
\ref{tbl:cmpError}).\\

Then, concerning the constriction area $A_{c}$, the mean quadratic error is
of 4.2 \% ($R^{2}_{A} = 0.71$). The theoretical and experimental curves
are depicted Figure \ref{img:cmp100-200} (b).
At the origin ($P_{e} = 0$ Pa), there is no significant error
between theory and experiments. 
This is not surprising since the values of $A_{c}$ for
$P_{e} = 0$ Pa and all the values of $P_{ext}$ have been used to determine
a Young modulus that gives the best adequation between simulations and data
(see \cite{chouly05a}). 
For values of $P_{e}$ between 0 Pa and 200 Pa ($0 \leq Re \leq 1000$), 
a plateau is observed in the experimental curves. 
It is explained by inertial effects in the deformable portion,
which are not taken into
account in the model because of the assumption of quasi-static deformation.
Nevertheless, it is not the major source of difference between the theoretical
and experimental curves.
For $200 \leq P_{e} \leq 400$ Pa
($1000 \leq Re \leq 1400$), $A_{c}$ is a decreasing function of $P_{e}$. 
This function, in a first approximation, should be treated as linear. 
In this range of values of $P_{e}$, the agreement between theory and
experiments can be considered as satisfying.
The maximal amplitude
of the closure ($\Delta A_{c}^{max} / A_{c}(P_{e} = 0)$) is of
approximatively 18 \%, 
and does not change significantly with $P_{ext}$. 
This maximal closure is always obtained for $P_{e} \simeq 400$ Pa.
Indeed, in the interval ($400 \leq P_{e} \leq 700$ Pa, $1400 \leq Re \leq 1900$), 
a change is observed:  the duct reopens at the level of the constriction. 
This behaviour might be
explained by  
an increase
of the downstream pressure $P_{c}^{16,exp}$ (up to $\simeq 180$ Pa,
after $P_{e} = 400$ Pa).
As $P_{c}^{16,exp}$
is used in the simulations to impose the values of the outlet pressure $P_{s}$, the numerical model
is able to reproduce this reopening, 
though the strong non-linearity
in the experimental curves 
is not reproduced in the simulations.
In the last interval ($P_{e} \geq 700$ Pa, $Re \geq 1900$), the theoretical and 
experimental curves are diverging. Indeed, after this critical value of $Re$,
self-sustained oscillations in the latex structure are initiated. No measurement
of $A_{c}$ is possible in this case. 
As the modelling assumptions include quasi-steadiness, such 
a behaviour can not be reproducted by the simulations.\\

\section{Discussion}
\label{sect:conclusions}

First, in the in-vitro experiments, flow-induced obstruction has been observed systema\-tically, with a ratio $\Delta A_{c}^{max} / A_{c}(P_{e} = 0)$ of
approximatively 18 \%. This
confirms the preliminary experimental results of \cite{choulyXXa}. This effect
is due to the internal airflow, which is submitted to acceleration at the level of the constriction
and induces pressure losses (Venturi effect). Since the deformation of the latex wall
is governed by the local pressure difference between the internal and external flow,
these pressure losses result in a decrease of the constriction area (partial obstruction
of the duct).
The values of the couple $(P_{e},P_{s})$ during an experiment are such that
in the  
quasi-steady regime ($200 \leq P_{e} \leq 400$ Pa), the expiratory phase 
is reproduced in a very simplified
manner. Remembering the analogy between the latex cylinder and the base of
the tongue, the airflow which circulates in the pipe is similar to an airflow
which would go from the hypopharynx to the mouth cavity. $P_{e}$ would then
be the pressure in the hypopharynx, at the base of the epiglottis, and $P_{s}$ the
pressure in the mouth cavity, approximatively equal to the atmospheric pressure
(as in the simplified models described in
\cite{lofaso98a,woodson03a}). Considering this analogy as valid,
the presented experiments would then reproduce the flow-induced obstruction
in conditions of expiratory flow, which has been 
observed in some apneic patients or heavy snorers, and may be at the origin
of expiratory flow limitation
\cite{martin80a,sanders83a,stanescu96a,woodson03a}.
This analogy, already dicussed in \cite{choulyXXa}, is however limited,
since the experiments do not reproduce with accuracy the complex dynamics of
a respiratory cycle (see e.g. \cite{yamashiro07a}), and was only 
focused on the quasi-steady phenomena.
The range of values of the upstream pressure $P_{e}$, however, is 
physiological as it conduces to airflow rates of the order of 10 l/min
(see e.g. \cite{trinder97a}). 
As a result, an extensive study of the response of the tongue replica
to different pressure commands, close to some typical physiological
or pathophysiological cases, may constitute a first perspective of this
work.
In particular, measurements in conditions closer to inspiration
($P_{e}$ equals to the atmospheric pressure, and $P_{s} < P_{e}$)
would allow to reproduce in a simplified manner what happens
during an apneic episode, which is known to occur during the inspiratory phase.
Of course, the change
from expiration to inspiration corresponds to a simple change of boundary conditions
in the numerical model, and simulations in inspiratory configuration have been carried
out with success from medical data (sagittal radiographies) \cite{chouly06b}.
Nevertheless, experiments in inspiratory configuration would be of
interest, on the one hand, to validate the simulations in this case, and on
the other, to obtain deformations of the latex structure of larger amplitudes. 
Furthermore, considering the actual geometry
of the setup, it is impossible to obtain a complete closure of the channel,
which would be of interest as Obstructive Sleep Apnea is associated to complete
closure of the pharynx and complete flow cessation.\\

Then, concerning the main point of this study, which is the experimental validation
of the numerical model described in Section \ref{sect:theory}, this one can be considered
to provide a satisfying first approximation for the prediction of the flow-induced collapse
measured with the in-vitro setup. 
Indeed, the main quadratic error for
the prediction of the constriction area is of  4.2 \% ($R^{2}_{A} = 0.71$).
This error does not vary significantly with
the pressure $P_{ext}$ in the external flow, when $P_{ext} \leq 500$ Pa.
For $P_{ext} = 700$ Pa, this error is higher and thus this value can be considered
as a limit of validity for the numerical model. 
Moreover, the theoretical and experimental curves are in good adequation for the
range of values of $P_{e}$ (200-400 Pa) corresponding to quasi-steady behaviour and thus relevant to our study.
When the upstream pressure is higher ($P_{e} \simeq 700$ 
Pa), auto-oscillations are observed in the experiments. 
This behaviour has already been
reported on symmetrical \cite{ruty07a} as well as on asymmetrical geometries
\cite{choulyXXa}. 
It can be predicted using linear stability theory in association with a simplified
physical model \cite{ruty07a}.
In our case and according to the assumption of quasi-steadiness, this phenomenon
can not be reproduced in the simulations. 
However, it should be associated to snoring or speech production
processes, and is therefore not relevant here.
The simplified asymptotic theory (RNSP) to model the fluid may be the first cause
of the differences between numerical predictions and measurements.
However, systematic comparisons effectuated with help of a rigid tongue replica
proved that this theory provides an accurate prediction of the pressure distribution for the
geometry we used, even if the assumption of small variations in the axial
direction necessary to derive the equations is not satisfied \cite{vanhirtum04a}.
From the earlier study \cite{vanhirtum04a}, it resulted also that the 
recirculation effects as well as turbulent
effects that may occur after separation of the flow had negligible impact on the
prediction of the pressure distribution. Moreover, no evidence of turbulence 
has been found in the region upstream the point of flow separation 
\cite{vanhirtum04a}.
The assumptions stated for the structure explain the other part of the
discrepancies between theory and experiments. In particular, the measurement
of the pressure at the level of the (initial) constriction ($x=0$ mm) revealed
that the numerical model underestimates the displacement of the tongue replica.
This might be due to the assumption of small displacements which is not
respected in this situation. An improvement would then consist in using
a shell or membrane theory with the assumption of large displacements
(geometrical non-linearity) for the finite element simulation. Yet, this
phenomenon does not have a strong effect on the final prediction
of the flow-induced obstruction.\\

The in-vitro setup allows to measure the pressure and the deformation 
during a fluid-structure interaction, in a configuration with strong asymmetry. 
Because of this asymmetry, the setup remains much more appropriate than other previous physical models, such as the collapsible tube 
(see e.g. \cite{ayappa03a,smith88a}), 
to study the human pharyngeal airway.
Furthermore, the measurement of the pressure at the level of the constriction,
which plays a critical role in the flow-induced collapse, coupled to the measurement
of the bidimensional geometry of the duct at the constriction, permitted to refine
the results and analysis carried out in \cite{choulyXXa}.
Nevertheless, as a simplification of the complex physiological reality, this
setup has some limitations.
For instance, it neglects the tridimensional effects that should be
involved in a true human airway, where lateral walls seem
to play a role during the collapse \cite{ayappa03a}.
In the protocol, it would also be interesting to measure the lateral deformation
of the tongue replica, so as to compare it to the numerical prediction.
More pressure sensors would also help to capture with more accuracy
the pressure distribution. 
\\

Finally, concerning the clinical implications of this work,
the numerical model proposed, as other models
based on continuum mechanics (see e.g. \cite{malhotra02a}),
is able to take into account easily and with 
relative accuracy patient-specific
geometrical and mechanical properties.
In particular, the choice of the Hooke law to model the soft tissue
is reasonable for patient-specific modelling, since
it appears difficult to obtain more information than the Young
modulus from in-vivo measurements. 
A second advantage of the proposed model 
is that the simulation time is reduced due to the simplificatory
assumptions.
It is of the order of 20 min on a
standard computer\footnote{Dell Precision 330 (TM) 
workstation, with Pentium(TM) 4, 2 GHz, 1 GO RAM.}, 
which may already be satisfying for a physiologist or a clinician.
With a simple optimization of the code, still in a preliminary
version, this time would be expected to be the order of one minute.
At last, the in-vitro validation described in this paper should
normally be an argument so that a clinician or a physiologist
may trust the predictions from the numerical model, though
in-vivo validation in this perspective is a must and remains
our ultimate goal.\\

\section{Conclusion}

Extensive experimental validation of a numerical model that predicts the flow-induced
collapse of the pharyngeal airway in conditions of strong obstruction has been carried
out. An in-vitro setup which reproduces the asymmetries and the particularities of the
airway at the base of the tongue in pathological conditions has been used for this
purpose. 
The prediction accuracy for the obstruction is of 4.2 \% (mean quadratic
error between prediction and measurements concerning the variation of the constriction
area), and has been assessed for a wide range of parameter values. 
Then, it results from these comparisons that the numerical model may be considered
as satisfying in a first approximation to predict the flow-induced deformation.
Since it is based on simplified assumptions, a low computational cost is associated
to each numerical simulation, which is an advantage for clinical applications.

\begin{acknowledgements}
The authors would like to thank
Pierre Chardon, Yves Garnier and
Freek van Uittert (Technische Universiteit Eindhoven, Netherlands)
for their very precious help on the in-vitro setup.
They would like also to thank Pr. Jean-Roch Paoli, Pr. Bernard Lacassagne
and Pr. Michel Tiberge (CHU Purpan, Toulouse, France) for their help on medical aspects.
\end{acknowledgements}

\nolinenumbers

\doublespacing

\bibliographystyle{spmpsci}
\begin{bibliography}{article}
\end{bibliography}

\linenumbers

\newpage


\newpage

\begin{table}[t]
\center

\vspace{7cm}

\begin{tabular}{p{30mm} p{15mm} p{15mm} p{15mm} p{15mm} p{15mm} p{15mm}}
$P_{ext} (Pa)$ & 100 & 200 & 300 & 500 & 700 & Mean \\ \hline 
$ P_{c,exp}^{0} (P_{e}^{max})$ (Pa) & 543 & 549 & 544 & 619 & 545 & 560\\ 
$ P_{c,num}^{0} (P_{e}^{max})$ (Pa) & 301 & 347 & 369 & 427 & 444 & 377\\ 
$\sigma_{exp} (P)$ (Pa) & 11.3 & 11.0 & 10.9 & 12.1 & 11.2 & 11.3 \\
\hline 
$\tilde{\varepsilon}_{2} (P)$ (\%)  & 11.2 & 9.8 & 9.2 & 9.2 & 9.9 & 9.8\\ 
$\tilde{\varepsilon}_{max} (P)$ (\%)  & 16.5 & 15.2 & 14.3 & 13.7 & 15.0 & 14.9 \\ \hline 
$R^{2}_{P}$ & 0.43 & 0.53 & 0.58 & 0.76 & 0.88 & 0.64 \\ 
\end{tabular}
(a)
\vspace{1cm}

\begin{tabular}{ p{30mm}  p{15mm} p{15mm} p{15mm} p{15mm} p{15mm} p{15mm}}
$P_{ext} (Pa)$ & 100 & 200 & 300 & 500 & 700 & Mean \\ \hline 
$ \Delta A_{c}^{max}$ ($\mathrm{mm^{2}}$) & 5.96 & 5.42 & 5.16 & 4.64 & 3.60 & 4.95 \\ 
$\sigma_{exp} (A)$ ($\mathrm{mm^{2}}$) & 0.26 & 0.30 & 0.29 & 0.27 & 0.31 & 0.29\\
\hline 
$\tilde{\varepsilon}_{2} (A)$ (\%)  & 3.2 & 3.1 & 3.2 & 4.1 & 7.2 & 4.2 \\ 
$\tilde{\varepsilon}_{max} (A)$ (\%)  & 6.0 & 6.1 & 6.5 & 7.4 & 12.6 & 7.7 \\ \hline 
$R^{2}_{A}$ & 0.78 & 0.83 & 0.72 & 0.67 & 0.53 & 0.71 \\
\end{tabular}
(b)
\vspace{1cm}

\caption{
Comparison between the predictions and the measurements of 
the pressure at the constriction $P_{c}^{0}$ (a) and the constriction area 
$A_{c}$ (b), for different values of $P_{ext}$. 
}
\label{tbl:cmpError}
\end{table}

\clearpage
\newpage


\clearpage
\newpage

\clearpage
\newpage

\end{document}